\documentstyle[prd,aps]{revtex}
\def\qq{$q\bar q$}

\def\X{X(3872)}
\def\DD{$D\bar D$}
\def\DD*{$D\bar D^*$}
\def\BB*{$B\bar B^*$}
\def\D*D*{$D^*\bar D^*$}
\def\B*B*{$B\bar B^*$}

\begin{document}
\input{psfig.sty}%

\title{Isospin breaking of the narrow charmonium state of Belle  at 3872 MeV as a deuson}

\author{Nils A. T\"ornqvist\thanks{\tt{e-mail: nils.tornqvist@helsinki.fi}} \\
Department of Physical Sciences, \\ University of Helsinki, POB
64, FIN--00014}

\maketitle

\begin{abstract}
The  narrow charmonium state near 3872 MeV
reported by the Belle collaboration and confirmed by CDF  lies almost exactly at the $D^0\bar D^{*0}$ threshold.
As was predicted many years ago it can be a $D\bar D^*$
deuteronlike meson-meson state called a deuson. If so, its spin-parity
should be  $1^{++}$, or possibily $0^{-+}$, and its isospin  predominantly 0.
An important decay mode should be via $D^0\bar D^{*0}$ to $D^0\bar D^0\pi^0$.
Large isospin breaking is expected
because of the isospin mass splitting between the neutral and charged $D$ ($D^*$) mesons.
Because of this large
isospin breaking the decay X(3872)$\to J/\psi \rho$ would be allowed, while $J/\psi \sigma $ would  be
forbidden by C-parity, as indicated by the present data.
\end{abstract}
\medskip

$PACS:$ 12.39.Pu; 14.40Aq; 14.40Gx; 14.40.Lb
\medskip

 Recently the Belle collaboration\cite{Belle0308029} reported a new
narrow charmonium state  at $3872\pm 0.6(stat) \pm 0.5(syst)$ MeV and with a width
$<2.3$ MeV (95\% C.L.), which is consistent with their resolution.
This new state, X(3782), has been confirmed by the CDFII collaboration at Fermilab\cite{CDFII}
(with a mass 3871.3$\pm0.7\pm0.4$ MeV). It is
60-100 MeV above the expected spin 2 $c\bar c $ ($^3D_{c2}$)
state\cite{Eichten80,Buchmuller}. Belle sees the  new state in the $\pi^+\pi^-J/\psi$
invariant mass distribution from $B^\pm$ decays produced in $e^+e^-$ collisions at the $\Upsilon(4)$ resonance.
They find  $34.1\pm 6.9\pm 4.1$ events of $B\to K^\pm
\pi^+\pi^-J/\psi$ and  a
$10.3\sigma$ signal significance   for the observed resonance peak.
CDF has 730$\pm$90 X(3872) candidates in the $J/\psi \pi^+\pi^-$ mass distribution
produced from initial $p\bar p$ collisions. Many recent papers have discussed possible quantum numbers and
interpretations of this new state\cite{close,pakvasa,voloshin,yuan,Wong,braathen,Barnes,swanson,Tomasz,bicudo,Eichten}
finding that interpreting the \X\ as a conventional $c\bar c$ state has difficulties. On the other hand, a multiquark or 2-meson composition, whose
existence have been suggested long ago\cite{band,volo,deruh} seems more likely. The fact that it is very narrow, excludes
many states which should decay strongly to $D\bar D$. Furthermore the \X\ mass is is not near any expected charmonium state
 and the nonobservation of $X \to \chi_{c1}\gamma$ suggests $C=+$. Therefore the \X\ may well be a non-\qq\ state.

Here we discuss the possibility that \X\
is a $D\bar D^*$ deuteron-like state  or a deuson. After Belle announced
their first note on their observation on the X(3872)\cite{Belle0308029} I wrote a short reminder\cite{reminder}, for the web, that such a state
had been predicted long ago with specific quantum numbers near the \DD*\ threshold together
with many others near the \D*D*\ thresholds and about 50 MeV below the
 \BB*\ and \B*B*\ thresholds. See Table 1.
 Here we elaborate on the possibility that \X\ is such a deuson
 and study in particular the large isospin breaking for such a state.


No doubt the deuteron is a multiquark state which to a good
approximation can be understood as a proton-neutron system bound
by mainly pion exchange. It is therefore natural to ask\cite{Tornqvist92} the question:
For which quantum numbers of two-meson composites is the well known pion
exchange mechanism attractive and comparatively large? This was
found to be the case for several light meson-meson channels  with
quantum numbers where problematic resonances have been seen, but the attraction
was found to be not strong enough to alone bind two light mesons such as $K\bar K^*$.
Then extending the
model to the heavy meson sector\cite{Tornqvist94}  using a similar
simple nonrelativistic
framework as for the deuteron\cite{Glen}  many deuteronlike
states were found to be bound or nearly bound. These are listed  in table 1. At that time similar arguments were presented by
Manohar and Wise\cite{Mano} and Ericson and Karl\cite{Karl}, although not with explicit model calculations.
\begin{table}[htb]
\begin{tabular}{l l l |l l l }
\hline
       Composite  & $J^{PC}$\ \ \ &     Mass [MeV] &  Composite  &
$J^{PC}$\ \ \ & Mass [MeV]
\\
\hline\hline $D\bar D^*$ & $0^{-+}$& $\approx 3870$ & $B\bar B^*$
& $0^{-+}$& $\approx
10545$            \\
      $D\bar D^*$ & $1^{++}$& $\approx 3870 $ &  $B\bar B^*$ &
$1^{++}$& $\approx
10562$                   \\
\hline $D^*\bar D^*$ & $0^{++}$& $\approx 4015 $ & $B^*\bar B^*$ &
$0^{++}$& $\approx
10582$                    \\
$D^*\bar D^*$ & $0^{-+}$& $\approx 4015 $ & $B^*\bar B^*$ &
$0^{-+}$& $\approx
10590$                   \\
$D^*\bar D^*$ & $1^{+-}$& $\approx 4015 $ &  $B^*\bar B^*$ &
$1^{+-}$& $\approx
10608$                    \\
$D^*\bar D^*$ & $2^{++}$& $\approx 4015 $ &  $B^*\bar B^*$ &
$2^{++}$& $\approx
10602$                 \\
\hline
\end{tabular}
\centering \caption{Predicted masses\protect\cite{Tornqvist94} of
heavy  deuteronlike states called deusons. These are close to the
$D\bar D^*$ and the $D^*\bar D^*$ thresholds, and about 50 MeV
below the $B\bar B^*$ and $B^*\bar B^*$ thresholds. All states
have I=0. The mass values were obtained from (a rather
conservative) one-pion exchange contribution only.}
\label{tab:heavydeusons}
\end{table}
The \X\ looks very much like one of  the two first deuteronlike
$D\bar D^*$ states predicted near 3870 MeV (see table 1 which is from table 8
of Ref.\cite{Tornqvist94}). The states are of course all
eigenstates of C-parity, or $(|D\bar D^*>\pm |\bar DD^*>)/\sqrt 2$
although we here denote them $D\bar D^*$ (and in the figures even without the "bar" $D D^*$).
 We note in particular:
\begin{itemize}
\item Its spin-parity should  be  $1^{++}$ or possibily $0^{-+}$.
For other quantum numbers pion exchange is
repulsive or so weak that bound states should not be expected.

\item No $D\bar D$ nor $B\bar B$
deusons are expected since the three pseudoscalar coupling (in
this case the $DD\pi$ coupling) vanishes because of parity.

\item If isospin were exact the Belle state as a deuson
would be a pure isosinglet with a mass very
close to the $D\bar D^*$ threshold. For  isovector states pion
exchange is generally one third weaker than for isoscalar states.
Therefore all predicted states were isosinglets. But as we shall see the
 isospin breaking will be large.

\item
The observed peak (at 3872 MeV) is within error bars at
the  $D^0\bar D^{*0}$ threshold (3871.2 MeV), while the
$D^+D^{*-}$ channel, 8.1 MeV higher,  is closed by phase space.

\item As a deuteronlike state with small binding energy (for
the deuteron it is 2.22 MeV) the Belle state should be large in
spatial size. It should then have a very narrow width since
annihilation of the loosely bound $D\bar D^*$ state to other
hadrons is expected to be small, although states containing the
$J/\psi$ are  favoured compared to states with only light hadrons
due to the OZI rule.

\item A large part of its width should be given by the
instability of its components or the $D^*$ widths, corrected for
phase space effects. An important decay should be via $D^0\bar
D^{*0}$ component to $D^0\bar D^0\pi^0$ since the other charge modes lie
about 2 MeV above the resonance. Using isospin  for the decay couplings
and the width measurements\cite{Anastassov02} for $D^{*+}$ together
with the dominance of the $D^0\bar D^{*0}$ component estimated below,
one finds that the width to $D^0\bar D^0\pi^0$ should be of the order 30-40 keV, and the
$D^0\bar D^0\gamma/D^0\bar D^0\pi^0$ ratio 62\% .
\end{itemize}

{\it The deuteron}

Let us now discuss our model for a deuteronlike \X\ in little more detail. We begin by recalling the essential
facts for understanding the deuteron. There pion exchange is the dominant mechanism which
determines the long range part of the proton-neutron wave function and most of the binding energy.
It is important to remember that, as for the deuteron, the potential
 comes from an axial dipole-dipole interaction where the tensor potential with its
 $1/r^3$ term ($T(r)$ below) is very large (although it must be regularized at short distances).
 Without the tensor potential and the associated D-wave the deuteron would not be bound. The cental
 potential $C(r)$ provides only about one third of the binding energy, while the freedom for the deuteron to flip the
 nucleon spins and make transitions between a $ ^3S_1$ and a $^3D_1$ state through the tensor force lowers the energy enough to
 be bound. This is important  in spite of the fact that the  D-wave interaction  is repulsive,
  and the fact that the D-wave has an angular momentum barrier.

The central and tensor parts of the potential are given by the functions
\begin{eqnarray}
C(r)&=& \frac   {e^{-m_\pi r}}{m_\pi r} \ ,\label{Cr} \\
T(r)&=& C(r) [1+\frac 3{m_\pi r} +\frac 3 {(m_\pi r)^2}] \ ,\label{Tr}
\end{eqnarray}
As can be seen from these expressions  for distances below a few fermi the tensor potential $T(r)$ is an order of magnitude larger than the central
$C(r)$, and can therefore not be neglected. In matrix form the deuteron potential can be writtem compactly in the form
\begin{eqnarray}
V_{deuteron}& =& -\frac{25}9 V_0  \left[
\left( \begin{array}{cc} 1&   0      \\    0     & 1\end{array}\right) C(r) +
\left( \begin{array}{cc} 0& \sqrt 8 \\ -\sqrt 8 & -2\end{array}\right) T(r)
\right]\ ,  \label{Vdeut}
\end{eqnarray}
where the first matrix index refers to the $^3S_1$ component  and the second index to $^3D_1$.
The factor $V_0=\frac{m_\pi^3g^2}{12\pi F^2_\pi}\approx 1.3$ MeV is fixed by the pion-nucleon coupling constant ($g$),
the pion decay constant, and by the constant 25/9 which
comes from the spin-isospin factor for the $^3S_1$ state.

Another fact one should remember is that one cannot predict exactly the deuteron binding energy,
without a very good detailed knowledge of the very short
range interactions. The actual measured value, 2.22457 MeV, is a net effect of different contributions of opposite signs,
most importantly from the repulsive kinetic term and the attraction from all the potential terms. One cannot
predict its value by scaling arguments from the other scales in the problem. But, given  the pion-nucleon coupling constant,
and a regularization model for the tensor potential,  one can say that
two nucleons is most likely to be bound, mainly by pion exchange, and with a small binging energy
for I=0 and for a specific mixture of  $^3S_1$ and $^3D_1$. For other quantum numbers the binding
is much weaker or repulsive, and no bound states exist, (although the $^1S_0$ state is nearly bound).

But once the binding energy is known and used as an input, to fix the regularization of the tensor potential at short distances,
the dominant long distance parts of the wave function of the deuteron can be predicted. See Fig. 1.
We use the same regularisation as in \cite{Tornqvist94} with a
parameter $\Lambda$, which can be related to a spherical
pion source around the hadron with rms radius $\sqrt{10}/\Lambda=0.624$fm/[$\Lambda$/GeV]. From the wave functions one can
then predict static properties of the deuteron such as the quadrupole moment in agreement with experiment.

\begin{figure}[]
\centerline{
\protect
\hbox{
\psfig{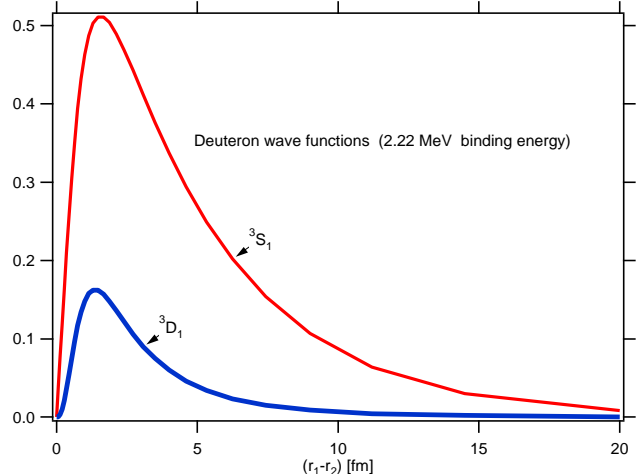}}}
\caption{The deuteron wave functions $^3S_1$ and $^3D_1$, when the the pion exchange potential is
regularized at small distances to get the right binding energy. Apart for very short distances this agrees with more detailed
models.}
\end{figure}

{\it The \X\ as a $D\bar D^*$ deuson}

This same argument as for the deuteron hold of course true also for deusons.
One can  determine for which quantum numbers pion exchange is attractive and strong
enough so that bound states are expected, but hardly the exact binding energy.
 But from the strength and sign of the potential one can exclude  already a large number of possible deuson candidates with given flavour, C-parity
and spin-parity quantum numbers. In fact only for $J^{PC}=1^{++}$ and possibily $0^{-+}$ and for I=0
(in the exact isospin limit) can one expect such states to exist just below the \DD*\ threshold, as was found in\cite{Tornqvist94}.

 For a $1^{++}$ $D\bar D^*$ system one has a rather similar potential as for the deuteron
 with an $^3S_1$ and a $^3D_1$-wave. It can be written in a similar form as
 Eq.(\ref{Vdeut}). One finds in the exact isospin limit a potential
\begin{eqnarray}
V_{1^{++}}& =& -\gamma V_0  \left[
\left( \begin{array}{cc} 1&   0      \\    0     & 1\end{array}\right) C(r) +
\left( \begin{array}{cc} 0& -\sqrt 2 \\ -\sqrt 2 & 1\end{array}\right) T(r)
\right]\ ,  \label{Vpvax}
\end{eqnarray}

\begin{figure}[]
\centerline{
\protect
\hbox{
\psfig{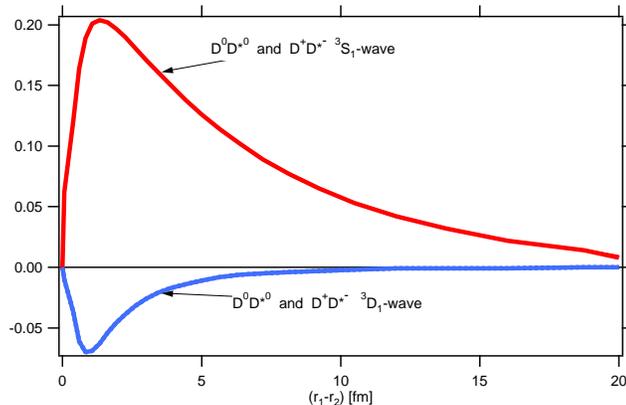}}}
\caption{The wave functions for a $1^{++}$ \DD*\ deuson assuming exact isospin invariance and 0.5 MeV binding energy.
(The  regularization parameter $\Lambda$ is here 1.355 GeV and the isospin splittings between charged and neutral $D,D^*$ and $\pi$
are neglected, by  using  the average $D,D^*,\pi$ masses in the calculation.)}
\end{figure}
\noindent where the overall constant $\gamma=3$ is determined by the spin-isospin and C-parity of the state\cite{Tornqvist94}.
One finds assuming a 0.5 MeV binding energy the wave functions shown in Fig.2.

{\it Isospin breaking}

However, since the binding energy of a such a deuteronlike state is expected to be at most a few MeV or
 of the same order as the isospin mass
splittings one must expect substantial isospin breaking in the wave functions.
 For a pure $I=0$ state one has equal contribution of the two components $D^ 0\bar D^{*0}$ and $D^+ D^{*-}$
in $(|D^ 0\bar D^{*0}>+|D^+ D^{*-}>)/\sqrt 2$, but since $D^ 0\bar
D^{*0}$ is 8.1 MeV lighter than $D^+ D^{*-}$ it should have a
greater weight than $D^+ D^{*-}$. In other words this means that there is a substantial $I=1$
component in the state. Now,
assuming exact isospin for the pion couplings, the $|D^ 0\bar D^{*0}>$ and the $|D^ +\bar D^{*-}>$ channels
are coupled through the matrix
\begin{eqnarray}
\left( \begin{array}{cc} 1& 2 \\ 2 & 1\end{array}\right)
\label{isomatrix}
\end{eqnarray}
The eigenvalues of this matrix are 3 and -1. The first corresponds to $\gamma=3$ in Eq.(\ref{Vpvax}) i.e. to the I=0 linear combination,
 where one has attraction, while the second (-1) corresponds to the weaker and repulsive I=1 \DD*\ channel.

Now to estimate the isospin breaking we
take account the fact that the  charge modes of the $D$ , $D^*$ and the pion are split in mass. Therefore ({\it a})
the kinetic terms for $|D^ 0\bar D^{*0}>$ and  $|D^ +\bar D^{*-}>$ are slightly different, ({\it b})
the potential is slightly different for
$\pi^0$ exchange than for $\pi^+$ echange, and ({\it c}) most importantly the two thresholds $|D^ 0\bar D^{*0}>$ and  $|D^ +\bar D^{*-}>$ are spit by 8.1 MeV.
We still assume that the couplings in the pion exchange are  related by exact isospin which gave the matrix (\ref{isomatrix}).
Clearly one expects that there will be only one bound state (i.e. if there is a bound state near $|D^ 0\bar D^{*0}>$
there should be no state near $|D^ +\bar D^{*-}>$ nor $|D^\pm\bar D^{*0}>$).
The perturbation from isospin breaking can only increase the binding of the lower state and make the orthogonal state
feel even more repulsion.

One then has a four coupled channels, two spin orbitals (S and D-wave)
coupled with two charge modes $|D^ 0\bar D^{*0}>$ and the $|D^ +\bar D^{*-}>$. The computed wave functions are shown in Fig. 3 when one assumes a
binding energy of half an MeV with respect to the lower threshold $|D^ 0\bar D^{*0}>$.
\begin{figure}[]
\centerline{
\protect
\hbox{
\psfig{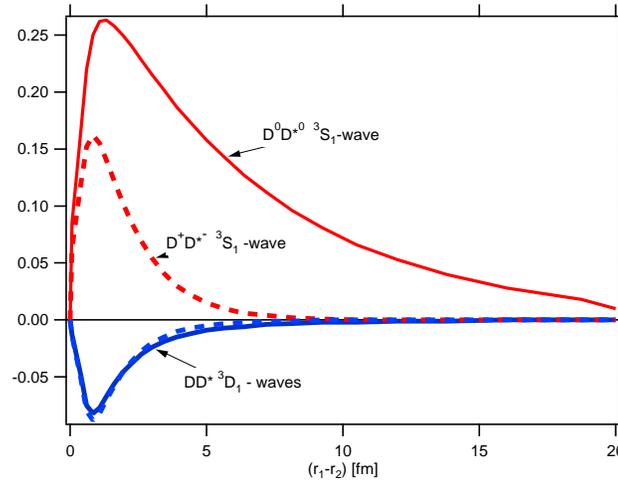}}}
\caption{ The 4 different components of the wavefunction for a $1^{++}$ \DD*\ deuson when isospin is brokren by the isospin mass splittings
of the $D$, $D^*$ and $\pi$. The binding energy with respect to the $D^ 0\bar D^{*0}$ threshold is assumed to be 0.5 MeV. Note the strong dominance of the
$|D^ 0\bar D^{*0}>$ (full lines) compared to the $|D^ +\bar D^{*-}>$ (dashed) in the S-wave.
(The  regularization parameter $\Lambda$ is here 1.40 GeV). }
\end{figure}

One notices that the $|D^ 0\bar D^{*0}>$ component in the wave function strongly
dominates over the $|D^ +\bar D^{*-}>$ in the S-waves for larger distances
as could be expected. Therefore in terms of isospin states there is a substantial I=1 component in the state.
As a consequence of this  the
decay of the \X\ to $\rho^0 J/\psi\to \pi^+\pi^-J/\psi$ will
not be forbidden. There are  indications of this in the
experiment\cite{Belle0308029}. On the other hand the decay chain
$\sigma J/\psi \to \pi^+\pi^-J/\psi$, where $\sigma$ is any
isoscalar object, is forbidden by  C-parity and spin-parity for a
$J^{PC}=0^{-+}$ or $1^{++}$ deuson. Thus one should expect some
$\rho^0 J/\psi \to \pi^+\pi^-J/\psi$, but no $\sigma J/\psi$ and in particular no $J/\psi \pi^0 \pi^0$. An experimental observation
of the $J/\psi \pi^0 \pi^0$ would exclude the \X\ to be a deuson. This is unfortunately a difficult mode to see experimentally and
it seems easier to with more statistics confirm that the $\pi^+\pi^-$ in $\pi^+\pi^-J/\psi$ comes only from $\rho$.

The second  case when pion exchange is attractive and strong is for a P-wave
(near) isoscalar $D\bar D^*$ system with $J^{PC}=0^{-+}$. The important thing here to notice is that the
large tensor term adds directly to the central term and not as in eq.(\ref{Vpvax}) through off-diagonal terms
and iteration. The large $1/r^3$ term can then in principle overcome  the P-wave angular momentum barrier!
The potential  is

\begin{eqnarray}
V_{0^{-+}}& =& \gamma V_0 [C(r)+2T(r)]=
-\gamma_{I\pm} V_0 \frac {\mu^2}{m_\pi^2}\frac{e^{-\mu r}}{m_\pi r} \left[ 3
 +\frac {6}{\mu r}+\frac {6}{(\mu r)^2} \right]\ , \label{Vpv0} \\
\end{eqnarray}
which  after a regularization can bind a pseudoscalar $D\bar D^*$ deuson in spite of the P-wave. The existence of this state
is however more model dependent than in the previous case since it is much more sensitive to the regularization of the tensor force, and
to other short range contributions.

Thus although the quantum numbers
$J^{PC}=1^{++}$ seems favoured $J^{PC}=0^{-+}$ is still not excluded that the \X\ is a pseudoscalar $D\bar D^*$ deuson.
The fact that in that case
the central and the large tensor potential adds directly to the same spin orbital
is similar to a two nucleon
system with $^3P_0$ quantum numbers, but with the important difference that one now has attraction, not repulsion from  pion exchange.
In Fig. 4 (assuming exact isospin) and Fig. 5 (including isospin breaking) we show the wave function for such a possible
pseudoscalar deuson when the binding energy is again set to be 0.5 MeV below the lower threshold.

\begin{figure}[]
\centerline{
\protect
\hbox{
\psfig{file=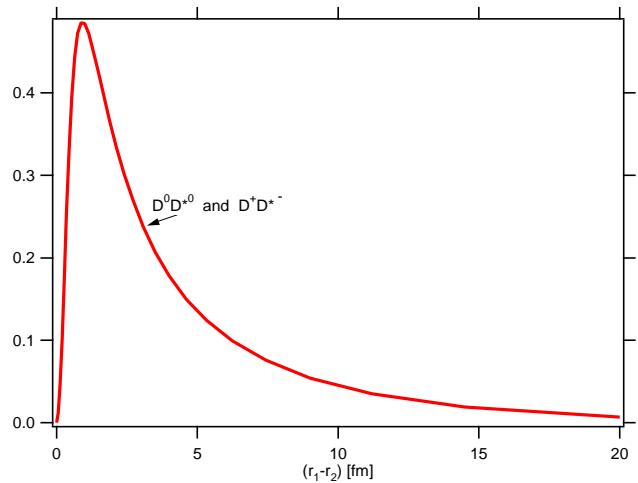,width= .50\textwidth,angle=0}}}
\caption{ The \DD*\ pseudoscalar deuson wave function  in the exact isospin limit, when the binding energy is assumed to be 0.5 MeV.
(The regularization parameter  $\Lambda$ is here 1.46 GeV).}
\end{figure}

\begin{figure}[]
\centerline{
\protect
\hbox{
\psfig{file=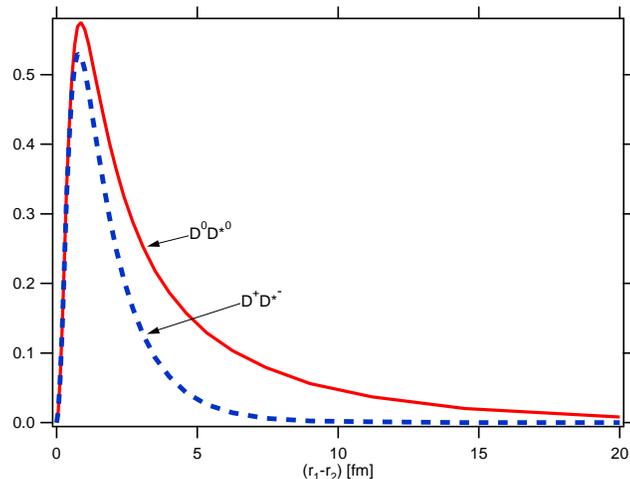,width= .50\textwidth,angle=0}}}
\caption{ The  $|D^ 0\bar D^{*0}>$ and  $|D^+\bar D^{*-}>$  pseudoscalar deuson
wave functions with isospin breaking and when the binding
energy below the $D^0D^{*0}$
threshols is 0.5 MeV.
(The regularization parameter $\Lambda$ is here 1.53 GeV.)}
\end{figure}

{\it Concluding remarks}

We conclude with a few general comments.
 The heavier the constituents are the stronger is the binding, since
the kinetic repulsion becomes smaller and is more easily overcome
by the attraction from the potential term. Thus as seen from table
1 the $D\bar D^*$ and $D^*\bar D^*$ systems are barely bound but
for $B\bar B^*$ and $B^*\bar B^*$ the binding energy is $\approx
50$ MeV \cite{Tornqvist94}.

An uncertainty in the original calculation\cite{Tornqvist94} was the $D^*$
coupling to $D\pi$, which was modelled from the $NN\pi$ coupling.
We predicted a $D^{*+}\to D^0\pi^+$ width of 63.3 keV in excellent
agreement with the recent measurement of $65\pm 3$
keV\cite{Anastassov02}. This increases the reliability of that
calculation, and we have here kept the same overall normalization $V_0=1.3$ MeV.

For flavour exotic
two-meson systems ($I=2$,  double strange, charm  or bottom), such
as $DD^*$ or $B^*B^*$, pion exchange is
always either too weakly attractive or repulsive.  Calculations do not
support such bound states to exist from pion exchange alone, and
shorter range forces are expected to be repulsive. Should flavour exotic $BB^*$ or $B^*B^*$
states exist, however (See \cite{Mano}), they would be quite
extremely narrow since they would be stable against strong decays.

In this calculation we have for simplicity not included the threshold effects from $\omega J/\psi$, $\rho^0 J/\psi$ nor  three body
channels $D\bar D\pi$ etc. into the coupled channel framework.
 In particular  $\omega J/\psi$ (which is just above the \X)
can be important. Swanson\cite{swanson} finds a rather large coupling of the \X\ to $\omega J/\psi$
when including a short range quark exchange potential. Then a substantial
$\omega J/\psi$ component can be generated into the wave function. This then  contributes to the binding of \X\
through higher order terms like $X \to \omega J/\psi \to X$. This mechanism is similar to the fact that the deuteron
binding increases because of the freedom for the $^3S_1$ state to turn into a $^3D_1$ and back.
It would  be important to have a measurement of the ratio of branching ratios,
${\cal B}(X\to D^0\bar D^0\pi^0)/{\cal B}(X\to J/\psi \pi\pi)$, which should not be too small for X as a deuson.
Belle\cite{BelleChistov} has measured an upper limit
for the product ${\cal B}(B^+\to XK^+)\times {\cal B}(X  \to D^0\bar D^0\pi^0)< 6\times 10^{-5}$, but this is not
yet significant because of the small ${\cal B}(B^+\to XK^+)$ compared to ${\cal B}(B^+\to \psi(3770)K^+)$.

Another recently seen peak where pion exchange should be
important is the observation of BES\cite{BES}  in $J/\psi\to \gamma
p\bar p$ near the $p\bar p$ threshold. Pion exchange should be the most
attractive for $N\bar N$ quantum numbers when I=0, L=0, J=0 or for
a pseudoscalar $^1S_0$ state. These are also the likely quantum numbers for the
peak which they have observed just above the $p\bar p$ threshold. However, if real, it need not necessarily be due
to a resonance below theshold, but could be due to a strong final state interaction
involving pion exchange between the produced proton-antiproton system\cite{Zou}.
Similar $p\bar p$ peaks but with low statistics have also been seen in $B^\pm\to p\bar p K^{\pm}$\cite{Belle1}  and
$\bar B^0\to p\bar p D^{(*)0}$\cite{Belle2}

 A more detailed understanding
with further experimental information on the Belle charmonium
state is important. If our arguments are supported by data it
could open up a completely new spectroscopy for heavy mesons.
It might also throw some
new light on many problematic light resonances in particular the $f_1(1420)$,
$\eta(1410)$ and  $\eta(1480)$ resonances just above the $K\bar
K^*$ threshold, where pion exchange in the final state interaction can play an important role,
although it is not expected to be strong enough to alone  bind those states. Possibly even the
glueball candidate $f_0(1500)$ could have a large $(\rho\rho -\omega\omega)/\sqrt 2$ component due to the
strongly attractive pion exchange for those quantum numbers\cite{Tornqvist92}.

{\it Acknowledgement}  I thank Steve Olsen and San Fu Tuan for useful
comments

\end{document}